# Distributions for Cited Articles from Individual Subjects and Years[1]

Mike Thelwall, Paul Wilson

The citations to a set of academic articles are typically unevenly shared, with many articles attracting few citations and few attracting many. It is important to know more precisely how citations are distributed in order to help statistical analyses of citations, especially for sets of articles from a single discipline and a small range of years, as normally used for research evaluation. This article fits discrete versions of the power law, the lognormal distribution and the hooked power law to 20 different Scopus categories, using citations to articles published in 2004 and ignoring uncited articles. The results show that, despite its popularity, the power law is not a suitable model for collections of articles from a single subject and year, even for the purpose of estimating the slope of the tail of the citation data. Both the hooked power law and the lognormal distributions fit best for some subjects but neither is a universal optimal choice and parameter estimates for both seem to be unreliable. Hence, only the hooked power law and discrete lognormal distributions should be considered for subject-and-year-based citation analysis in future and parameter estimates should always be interpreted cautiously.

## 1. Introduction

Citation counts are known to be highly skewed in the sense that, after a few years, many articles in a typical collection will have received few citations and a few articles will have received many citations (Price, 1965). Other phenomena also exhibit similar behaviour. For example, web hyperlink counts are also highly skewed, with a small number of pages attracting huge numbers of hyperlinks whereas many pages attract few (Barabási, & Albert, 1999). A mathematical distribution that has the same property, and which has been argued to model citations, either for all papers with at least one citation or for all papers with at least a moderate number of citations, is the power law (e.g., Albarrán, Crespo, Ortuño, & Ruiz-Castillo, 2011; Clauset, Shalizi, & Newman, 2001, 2009; Redner, 1998; Wen & Hsieh, 2013). This is the simple formula $1/x^\alpha$ with one parameter, $\alpha$. For example, if a power law is applied to the number of citations to a set of articles, then for some fixed value $\alpha$ it would be true that the probability that a randomly selected article had received $x$ citations would be proportional to $1/x^\alpha$. Nevertheless, other mathematical distributions have also been proposed, such as a discrete version of the lognormal distribution (Radicchi & Castellano, 2012) with continuous probability density function $\frac{1}{x\sqrt{2\pi}\sigma}e^{-\frac{(lnx-\mu)^2}{2\sigma^2}}$, where $\mu$ and $\sigma$ are its two parameters. For other types of data, investigations have shown that, despite initial claims, power laws are less appropriate than the lognormal distribution (e.g., Downey, 2001; Mitzenmacher, 2004) and so it is important to assess whether power laws, or variants, are appropriate for citation distributions.

---

[1] Thelwall, M. & Wilson, P. (2014). Distributions for cited articles from individual subjects and years. Journal of Informetrics, 8(4), 824-839.

An alternative possible distribution, based on the formula $1/(B+x)^\alpha$, where B > -1, and called here the *hooked power law*, is an extension of the power law where $B$ is a second parameter. This distribution was derived for web links (Pennock, Flake, Lawrence, et al., 2002 – see below) but seems to have not been used for citations although it is a logical alternative distribution, given the acknowledged parallels between hyperlinks and citations. This article assesses whether the hooked power law is a better fit than the lognormal and power law distributions for citation data for a single year and subject. Although a previous study has suggested that citation distributions from all fields can be scaled to give the same shape (Radicchi, Fortunato, & Castellano, 2008), this has subsequently been shown not to be the case (Waltman, van Eck, & van Raan, 2012) and so the issue is still unresolved. Moreover, a citation distribution that fits for long periods of time will not necessarily fit time-sliced data, such as that for all citations within a given year. Scientometric evaluations that use citations often compare articles separately that are published within a single year or within a short time window (e.g., for the main UK research evaluation, expert panels consider articles from several years but, if they use citations, are given field-based citation averages separately for each year under consideration) and so it is important to identify distributions that fit the type of data that is used in practice.

The question of which statistical model best fits citations is an abstract mathematical one, but the answer has both theoretical and practical implications (e.g., Glanzel, 2007; Ruiz-Castillo, 2013). It has practical implications for the use of citation analysis in research evaluation because power laws can be generated by feedback loops, and this appears to be the most common explanation for natural phenomena that obey power laws. In bibliometrics, the feedback loop is known as the Matthew effect (Merton, 1968) and elsewhere it is known as rich-get-richer (e.g., Adamic & Huberman, 2000). In other words, if citations obey a power law then, once cited, articles generate citations primarily because of their existing citations rather than because of their intrinsic value. If true, then this would be an argument against using citations in research evaluation. In contrast, a hooked power law is consistent with a process in which articles attract new citations partly due to their existing citations and partly due to other factors (Pennock, Flake, Lawrence, Glover, & Giles, 2002), such as their intrinsic worth. This would be more consistent with citations being used in research evaluation but would nevertheless be an argument against counting the citations to particularly highly cited articles (i.e., citation classics: Garfield, 1987) at face value because a high proportion may be due to existing citations rather than the intrinsic worth of the article. In bibliometrics, this may be related to the previously observed phenomena of the perfunctory citations given to works that appear to be merely concept markers (Case & Higgins, 2000). Although the hooked power law behaves like a power law when $x$ is much greater than $B$, the additional parameter $B$ suggests that another factor is also at work (see Appendix A).

It is also important to assess which distributions fit citation data best because this can help when conducting theoretical studies of factors that affect citation counts, such as the role of collaboration and countries (Didegah, & Thelwall, 2013; Peters, & van Raan, 1994) as well as when using citation counts for research evaluation purposes – typically for articles within a single subject and a single year or small time window. Such studies can potentially be more powerful if they use statistical methods that are best tailored to the appropriate citation distribution.

## 2. Research questions

The following questions address the goal of deciding which mathematical distribution is the most appropriate for citations from a single subject and year, and whether the distributions are suitable.

- Which out of the power law, the hooked power law, and the (discrete) lognormal distribution is the best model for the distribution of counts of citations to articles in individual subjects and years?
- In the above case, is the power law exponent extracted from an optimal fit similar to the corresponding hooked power law exponent?
- In the above case, can the parameters for the best fitting distributions be estimated to a high degree of precision?

## 3. Methods

In order to test the different distributions for the first two research questions, twenty different subject areas were selected from Scopus in order to cover a wide range of different potential citation norms. This is important because different areas of scholarship cite at different speeds and with different goals (Amin & Mabe, 2000). The subject areas include some from the arts, humanities, social sciences, physical sciences, formal sciences and medicine. Scopus was chosen in preference to the Web of Science for its larger coverage (de Moya-Anegón, Chinchilla-Rodríguez, Vargas-Quesada, et al., 2007) but this choice seems unlikely to affect the results.

     In order to give approximately ten years to attract citations, which should be enough to give substantial numbers of citations even in the slowest moving fields, articles were selected from the end of 2004, up to a maximum of 5000 (a Scopus search interface limitation). Due to the Scopus search interface the data sets included some articles from very early 2005 (typically 1 January 2005) but this should not bias the citation distributions extracted much. The most significant difference between the subjects is that some include a full year of articles (e.g., Algebra and Number Theory) whereas others contain only articles from later in 2004 (e.g., Biochemistry) but this should also not bias the results much because of the long time period (over ten years) involved (Table 1). For each subject the number of citations to each article (excluding reviews) was extracted from Scopus.

**Table 1**. The 20 Scopus subjects selected and the number of articles (excluding reviews and documents that are not articles) extracted from them from the end of 2004 (maximum 5000 articles). The data, including citations, was collected in May 2014.

| Scopus subject | Abbreviation | Articles | General subject area |
|---|---|---|---|
| Accounting | Accounting | 1178 | Business, Management and Accounting |
| Algebra and Number Theory | Algebra | 528 | Mathematics |
| Applied Mathematics | AppliedMaths | 5000 | Mathematics |
| Biochemistry | Biochem | 5000 | Biochemistry, Genetics and Molecular Biology |
| Dermatology | Dermatology | 3184 | Medicine |
| Developmental Biology | Developmental | 4541 | Biochemistry, Genetics and Molecular Biology |
| Ecology, Evolution, Behavior and Systematics | Ecology | 5000 | Agricultural and Biological Sciences |
| Genetics | Genetics | 5000 | Biochemistry, Genetics and Molecular Biology |
| History | History | 5000 | Arts and Humanities |
| Horticulture | Horticulture | 3009 | Agricultural and Biological Sciences |
| Literature and Literary Theory | Literature | 5000 | Arts and Humanities |
| Logic | Logic | 4547 | Mathematics |
| Marketing | Marketing | 1550 | Business, Management and Accounting |
| Oncology | Oncology | 4646 | Medicine |
| Rehabilitation | Rehab | 5000 | Medicine |
| Soil Science | Soil | 4347 | Agricultural and Biological Sciences |
| Statistics and Probability | StatsProb | 5000 | Mathematics |
| Tourism, Leisure and Hospitality Management | Tourism | 608 | Business, Management and Accounting |
| Urology | Urology | 5000 | Medicine |
| Visual Arts and Performing Arts | Visual | 4096 | Arts and Humanities |

When a statistical distribution is believed to be suitable for given data, the first step for verifying this is to estimate the parameters of the distribution. Statistical distributions are often fitted to data according to the maximum likelihood principal (ordinary least squares regression is also common for some contexts). This asserts that a distribution should be fitted to data by choosing values of parameters that maximise the likelihood that the data was derived from the distribution. For any set of parameters (e.g., $\alpha$, $B$ in the case of the hooked power law) this means calculating the probability that each data point could have been derived from the formula for those parameters, and then multiplying the probabilities together for all of the data points. Because this usually generates very small numbers, the log of this figure is used, which is called the log-likelihood of the parameters. The distribution is then fitted to the data by developing an algorithm to find values of the parameters that maximise the log-likelihood (or minimise it if, as in the current paper, the log-likelihood values are multiplied by -1 for convenience). The distributions in this paper were fitted using an analytical process (based on derivatives) for the power law and the lognormal distribution and a simple iterative gradient descent algorithm for the hooked power law (see Appendix B).

Data sets often have different behaviours at zero or small numbers than for large numbers. In response, zero-inflated models have been developed to help analyse datasets where the zeroes are distributed differently to the rest of the numbers (e.g., Hilbe, 2011). Similarly, power laws are often fitted to only the tails of data sets, ignoring all points less than a specified minimum value in the belief that low numbers may be driven by different processes than are high numbers (Clauset, Shalizi, & Newman, 2009). In terms of citations, this would mean picking a number, $x_{min}$, and ignoring all articles with less than $x_{min}$ citations when fitting the distribution. The standard procedure for this, which is used here, is to try different values of $x_{min}$ and fit the distribution for each value, keeping the $x_{min}$ for which the best fit can be obtained (Clauset, Shalizi, & Newman, 2009).

The three candidate distributions (discrete versions of the power law, lognormal and hooked power law) were fitted to the data and compared against each other under three conditions and for each subject.

- *Optimal conditions for the power law*: For each subject, the minimum number of citations was set to give the power law the best fit to the data and then the other two distributions were also fitted with this same minimum.
- *Optimal conditions for the lognormal*: For each subject, the minimum number of citations was set to give the lognormal distribution the best fit to the data and then the other two distributions were also fitted with this same minimum.
- *All cited articles*: For each subject, the minimum number of citations was set to 1 to include all the citation data and then all three distributions were fitted with this minimum.

All three of the distributions are continuous but fitted here to discrete data. One way to perform this conversion, which is used here, is to use the continuous Probability Density Function (PDF) as the basis of the probability mass function for the discrete case. This can cause a problem because, whilst the integral of the continuous PDF is 1, the sum of all integer values of the PDF might not be 1, in which case it would not be a valid probability mass function. To correct for this issue, the continuous PDFs were converted to discrete probability mass functions by dividing them by their sum from one to infinity (or an approximation) of the original point estimates. For example, if $f(x)$ is the lognormal PDF, so that $\int_0^\infty f(x) = 1$, then the point mass function (evaluated only at positive integer x) is $f(x)/\sum_{i=1}^\infty f(i)$. See Appendix B for specific details for the three distributions used.

The Vuong (1989) test is typically used to compare distributions. In the case of different types of distributions (non-nested distributions) the Vuong statistic is normally distributed and is calculated by comparing the log-likelihoods of the distributions, correcting for the sample size, the differing number of free parameters in each distribution, and the variability of the data. The Vuong test was used to compare the lognormal distribution against the power law and the lognormal distribution against the hooked power law. The power law distribution was compared against the hooked power law distribution with the Likelihood Ratio Test (LRT) (Wilks, 1938) because the former is nested within the latter (by setting B=0). The LRT test statistic is double the difference in the log-likelihoods, which follows the chi-square distribution with 1 degree of freedom (for the extra parameter *B*), giving critical values of 3.841 for p=0.05 and 6.635 for p=0.01.

For the third research question, simulations were run on the hooked power law and lognormal distributions in order to assess the precision of the parameter estimates.

# 4. Results

When fitting all three distributions to the data with the optimal minimum threshold for citations set for the power law (Table 2), in almost all cases all three distributions fit the data equally well. The exceptions are that the lognormal fits Developmental significantly better than does the power law and the hooked power law fits both History and Urology significantly better than does the lognormal distribution. The hooked power law also fits four subjects significantly better than does the power law. Moreover, for 19 of the 20 subjects the Vuong statistic for comparing the lognormal and the power law is negative, suggesting (but not giving statistical evidence) that the lognormal is a marginally better fit. Similarly, in about half of the cases the hooked power law is a better fit than the lognormal distribution, suggesting that the two are broadly equivalent for this data. Taken together, it seems reasonable to claim that although most of the time the three distributions fit the tail of the citation distributions equally well, the lognormal and hooked power law are both marginally better fits than the power law, even for the length of tail that the power law fits best.

When fitting all three distributions to the data with the optimal minimum threshold for citations set for the lognormal distribution (Table 3), the minimum citation thresholds are typically lower than for the power law (i.e., the distributions are less truncated). In all cases the hooked power law and lognormal distributions both fit the data significantly better than does the power law, showing that the lognormal distribution is consistent with citation counts over a greater range of their values. Overall, the hooked power law fits the data for this range about as well as the lognormal distribution and hence it also fits citation distributions over a greater range than does the power law.

When fitting all three distributions to all of the articles with at least one citation (Table 4), the hooked power law and lognormal distributions fit statistically significantly better than does the power law. The lognormal distribution fits statistically significantly better than does the hooked power law in only three subjects, whereas the hooked power law fits better for nine subjects. Whilst this suggests a slight preference for the hooked power law overall, the statistically significant results in both directions give confirmation that citations do not necessarily follow the same distribution or laws in all subjects. Figures 1 and 2 shows a case where the hooked power law fits better than the lognormal distribution and Figure 3 shows a reverse case.

**Table 2**. Comparison of the power law, hooked power law and lognormal distributions with the optimal minimum number of citations for the power law. Articles are taken from the first up to 5000 articles reported in Scopus in each subject for the year 2004.*

| Subject | Min. cit. | Articles | Pl alpha | LN mean | Ln SD | Hooked alpha | Hooked B | LL pl | LL ln | LL hook | Vuong (pl-ln) | Vuong (ln-h) | LRT (hook-pl) |
|---|---|---|---|---|---|---|---|---|---|---|---|---|---|
| Accounting | 59 | 146 | 2.982 | 3.6 | 0.81 | 7.620 | 212.9 | 713.89 | 710.29 | 710.22 | -1.628 | -0.196 | **7.3** |
| Algebra | 12 | 73 | 3.152 | 0.8 | 1.06 | 4.609 | 11.6 | 229.52 | 228.87 | 228.86 | -0.604 | -0.158 | 1.3 |
| AppliedMaths | 58 | 154 | 3.062 | -28.9 | 4.06 | 3.144 | 3.4 | 741.25 | 741.24 | 741.24 | -0.093 | -0.034 | 0.0 |
| Biochem | 58 | 487 | 3.152 | -640.4 | 17.33 | 3.152 | 0.0 | 2313.27 | 2313.30 | 2313.27 | 0.793 | -0.816 | 0.0 |
| Dermatology | 37 | 119 | 3.538 | 1.2 | 1.11 | 4.720 | 23.9 | 483.21 | 482.66 | 482.75 | -0.702 | 1.195 | 0.9 |
| Developmental | 55 | 761 | 2.671 | 2.4 | 1.26 | 3.672 | 53.4 | 3868.13 | 3859.46 | 3860.29 | **-2.517** | 1.600 | **15.7** |
| Ecology | 86 | 201 | 3.667 | -1.4 | 1.57 | 4.162 | 21.9 | 973.37 | 973.13 | 973.18 | -0.414 | 0.579 | 0.4 |
| Genetics | 118 | 316 | 2.822 | -4.9 | 2.42 | 3.072 | 25.4 | 1806.00 | 1805.75 | 1805.68 | -0.376 | -0.296 | 0.6 |
| History | 5 | 824 | 2.238 | 0.7 | 1.31 | 3.229 | 6.9 | 2560.85 | 2541.14 | 2542.87 | -3.819 | **2.181** | **36.0** |
| Horticulture | 38 | 327 | 3.307 | 0.5 | 1.31 | 4.276 | 22.8 | 1380.70 | 1379.50 | 1379.17 | -0.666 | -1.024 | 3.1 |
| Literature | 9 | 51 | 4.129 | -0.3 | 0.98 | 5.653 | 5.5 | 118.77 | 118.59 | 118.64 | -0.418 | 1.136 | 0.3 |
| Logic | 58 | 176 | 3.765 | -2.2 | 1.59 | 4.184 | 11.9 | 773.76 | 773.58 | 773.66 | -0.397 | 0.941 | 0.2 |
| Marketing | 47 | 247 | 2.882 | -13.4 | 3.11 | 3.081 | 7.6 | 1170.51 | 1170.42 | 1170.35 | -0.189 | -0.374 | 0.3 |
| Oncology | 88 | 486 | 2.754 | -11.3 | 3.10 | 2.899 | 11.7 | 2663.33 | 2663.17 | 2663.11 | -0.346 | -0.294 | 0.4 |
| Rehab | 49 | 208 | 3.348 | 2.4 | 0.97 | 5.256 | 57.0 | 926.44 | 924.57 | 924.69 | -1.092 | 0.759 | 3.5 |
| Soil | 63 | 184 | 3.831 | 0.5 | 1.23 | 4.673 | 25.3 | 818.41 | 817.99 | 818.09 | -0.595 | 1.408 | 0.6 |
| StatsProb | 79 | 159 | 2.692 | -11.2 | 3.14 | 2.792 | 7.6 | 863.03 | 862.98 | 862.99 | -0.237 | 0.154 | 0.1 |
| Tourism | 32 | 127 | 2.846 | 3.0 | 0.84 | 6.997 | 113.4 | 556.13 | 552.66 | 552.64 | -1.673 | -0.055 | **7.0** |
| Urology | 58 | 380 | 3.172 | 1.7 | 1.21 | 4.174 | 39.2 | 1799.82 | 1797.61 | 1797.99 | -1.350 | **2.113** | 3.7 |
| Visual | 5 | 144 | 3.216 | -608.8 | 16.66 | 3.216 | 0.0 | 313.92 | 313.93 | 313.92 | -0.042 | -0.107 | 0.0 |

* pl = power law, ln = lognormal, hooked = hooked power law, LL = minus log likelihood – lower values indicate a better fit; Vuong = Vuong z values – bold if significant at p=0.05, where positive values indicate that the first bracketed model is a better fit than the second.

**Table 3**. Comparison of the power law, hooked power law and lognormal distributions with the optimal minimum number of citations for the lognormal distribution. Articles are taken from the first up to 5000 articles reported in Scopus in each subject for the year 2004.*

| Subject | Min. cit. | Articles | Pl alpha | LN mean | Ln SD | Hooked alpha | Hooked B | LL pl | LL ln | LL hook | Vuong (pl-ln) | Vuong (ln-h) | LRT (hook-pl) |
|---|---|---|---|---|---|---|---|---|---|---|---|---|---|
| Accounting | 5 | 865 | 1.637 | 2.8 | 1.13 | 3.956 | 55.8 | 3912.554 | 3766.5 | 3766.3 | **-10.892** | -0.198 | **292.5** |
| Algebra | 1 | 455 | 1.519 | 1.4 | 1.01 | 6.064 | 24.4 | 1413.146 | 1268.9 | 1270.4 | **-10.740** | 0.661 | **285.5** |
| AppliedMaths | 3 | 3180 | 1.686 | 2.0 | 1.15 | 3.495 | 20.6 | 11979.6 | 11534.6 | 11534.0 | **-17.766** | -0.316 | **891.2** |
| Biochem | 8 | 3751 | 1.835 | 3.0 | 0.92 | 5.212 | 77.5 | 16479.15 | 15913.6 | 15914.5 | **-16.189** | 0.133 | **1129.3** |
| Dermatology | 3 | 1941 | 1.818 | 1.7 | 1.10 | 3.834 | 16.4 | 6520.661 | 6313.6 | 6314.7 | **-12.459** | 1.552 | **411.9** |
| Developmental | 19 | 2409 | 2.156 | 2.7 | 1.19 | 3.455 | 41.4 | 11173.66 | 11083.9 | 11085.0 | **-7.965** | 0.838 | **177.3** |
| Ecology | 7 | 3788 | 1.767 | 3.0 | 0.84 | 7.857 | 143.4 | 16825.45 | 15994.2 | 16026.8 | **-24.945** | 4.977 | **1597.3** |
| Genetics | 44 | 1238 | 2.435 | 1.6 | 1.53 | 3.113 | 35.8 | 6324.049 | 6312.1 | 6311.8 | **-2.530** | -0.229 | **24.5** |
| History | 2 | 1774 | 1.871 | 0.8 | 1.29 | 2.994 | 5.0 | 4872.115 | 4766.6 | 4769.0 | **-8.782** | 1.588 | **206.2** |
| Horticulture | 13 | 1260 | 2.288 | 2.8 | 0.84 | 5.953 | 67.3 | 5103.5 | 5016.3 | 5016.4 | **-6.958** | 0.117 | **174.2** |
| Literature | 1 | 1652 | 1.983 | 0.2 | 0.99 | 4.742 | 4.5 | 2754.194 | 2632.8 | 2632.8 | **-9.208** | -0.070 | **242.8** |
| Logic | 15 | 1293 | 2.345 | 2.9 | 0.84 | 6.295 | 78.9 | 5330.786 | 5248.6 | 5249.0 | **-7.510** | 0.646 | **163.6** |
| Marketing | 2 | 1368 | 1.443 | 2.6 | 1.25 | 3.465 | 39.3 | 6144.25 | 5767.0 | 5763.9 | **-17.678** | -1.330 | **760.7** |
| Oncology | 6 | 3526 | 1.590 | 3.2 | 1.13 | 3.671 | 70.5 | 17340.97 | 16654.9 | 16651.5 | **-22.212** | -1.201 | **1378.9** |
| Rehab | 22 | 714 | 2.649 | 2.6 | 0.95 | 4.898 | 48.7 | 2980.591 | 2961.6 | 2961.8 | **-3.663** | 0.684 | **37.6** |
| Soil | 14 | 1711 | 2.292 | 2.8 | 0.85 | 6.308 | 79.7 | 7052.42 | 6936.9 | 6936.9 | **-8.981** | 0.024 | **231.0** |
| StatsProb | 2 | 3987 | 1.523 | 2.1 | 1.26 | 3.111 | 19.9 | 15923.78 | 15119.9 | 15112.4 | **-23.931** | **-2.520** | **1622.8** |
| Tourism | 7 | 429 | 1.836 | 2.8 | 0.93 | 6.085 | 86.2 | 1822.283 | 1758.9 | 1759.1 | **-7.155** | 0.234 | **126.4** |
| Urology | 12 | 2214 | 2.034 | 2.9 | 0.95 | 4.968 | 71.2 | 9690.694 | 9495.1 | 9495.3 | **-11.703** | 0.258 | **390.8** |
| Visual | 1 | 982 | 1.924 | 0.2 | 1.06 | 4.103 | 3.9 | 1749.381 | 1677.2 | 1675.8 | **-6.144** | -1.160 | **147.2** |

* pl = power law, ln = lognormal, hooked = hooked power law, LL = minus log likelihood – lower values indicate a better fit; Vuong = Vuong z values – bold if significant at p=0.05, where positive values indicate that the first bracketed model is a better fit than the second.

Table 4. Comparison of the power law, hooked power law and lognormal distributions for all articles with at least one citation. Articles are taken from the first up to 5000 articles reported in Scopus in each subject for the year 2004.*

| Subject | Min. cit. | Articles | Pl alpha | LN mean | Ln SD | Hooked alpha | Hook B | LL pl | LL ln | LL hook | Vuong (pl-ln) | Vuong (ln-h) | LRT (hook-pl) |
|---|---|---|---|---|---|---|---|---|---|---|---|---|---|
| Accounting | 1 | 1089 | 1.320 | 2.6 | 1.30 | 3.728 | 48.4 | 5131.4 | 4637.8 | 4631.9 | **-22.112** | 1.493 | **987.2** |
| Algebra | 1 | 455 | 1.519 | 1.4 | 1.01 | 6.064 | 24.4 | 1413.1 | 1268.9 | 1270.4 | **-10.740** | -0.661 | **288.4** |
| AppliedMaths | 1 | 4111 | 1.413 | 1.8 | 1.30 | 3.289 | 17.4 | 15605.5 | 14387.3 | 14373.4 | **-30.917** | **3.284** | **2436.4** |
| Biochem | 1 | 4811 | 1.302 | 2.8 | 1.07 | 6.848 | 130.0 | 23754.0 | 20579.4 | 20574.5 | **-54.364** | 0.260 | **6349.2** |
| Dermatology | 1 | 2687 | 1.465 | 1.6 | 1.18 | 3.816 | 16.1 | 9196.3 | 8420.4 | 8421.2 | **-25.146** | -0.253 | **1551.8** |
| Developmental | 1 | 4480 | 1.285 | 3.0 | 1.11 | 5.024 | 105.1 | 23183.3 | 20171.2 | 20233.2 | **-58.542** | **-4.010** | **6024.2** |
| Ecology | 1 | 4654 | 1.302 | 2.8 | 1.09 | 10.914 | 233.7 | 22961.5 | 19953.7 | 19841.8 | **-51.523** | **5.589** | **6015.6** |
| Genetics | 1 | 4710 | 1.290 | 2.9 | 1.32 | 3.525 | 61.0 | 24019.5 | 21661.9 | 21611.0 | **-47.320** | **5.569** | **4715.2** |
| History | 1 | 2637 | 1.646 | 0.6 | 1.36 | 2.769 | 3.7 | 6703.2 | 6434.6 | 6440.2 | **-13.955** | **-2.549** | **537.2** |
| Horticulture | 1 | 2820 | 1.356 | 2.3 | 1.16 | 6.118 | 70.3 | 12151.3 | 10825.6 | 10782.1 | **-34.037** | **4.732** | **2651.4** |
| Literature | 1 | 1652 | 1.983 | 0.2 | 0.99 | 4.742 | 4.5 | 2754.2 | 2632.8 | 2632.8 | **-9.208** | 0.070 | **242.8** |
| Logic | 1 | 3755 | 1.381 | 2.1 | 1.28 | 4.107 | 33.2 | 15273.0 | 13962.5 | 13918.6 | **-32.735** | **5.890** | **2621.0** |
| Marketing | 1 | 1459 | 1.328 | 2.5 | 1.32 | 3.437 | 38.5 | 6723.6 | 6112.2 | 6104.5 | **-23.784** | 1.929 | **1222.8** |
| Oncology | 1 | 4226 | 1.285 | 3.0 | 1.32 | 3.664 | 70.3 | 21902.3 | 19735.1 | 19670.6 | **-44.781** | **6.715** | **4334.4** |
| Rehab | 1 | 2904 | 1.415 | 1.6 | 1.67 | 2.514 | 10.3 | 10981.5 | 10485.3 | 10540.2 | **-19.362** | **-5.799** | **992.4** |
| Soil | 1 | 3961 | 1.352 | 2.3 | 1.18 | 5.873 | 69.5 | 17231.1 | 15397.8 | 15332.9 | **-40.102** | **6.317** | **3666.6** |
| StatsProb | 1 | 4455 | 1.382 | 2.0 | 1.33 | 3.056 | 18.8 | 18061.4 | 16626.7 | 16611.0 | **-33.728** | **3.729** | **2869.4** |
| Tourism | 1 | 593 | 1.327 | 2.5 | 1.13 | 6.812 | 103.4 | 2738.3 | 2409.5 | 2404.9 | **-17.965** | 0.963 | **657.6** |
| Urology | 1 | 4250 | 1.343 | 2.4 | 1.35 | 3.900 | 42.7 | 18873.9 | 17263.2 | 17196.6 | **-36.482** | **7.146** | **3221.4** |
| Visual | 1 | 982 | 1.924 | 0.2 | 1.06 | 4.103 | 3.9 | 1749.4 | 1677.2 | 1675.8 | **-6.144** | 1.160 | **144.4** |

* pl = power law, ln = lognormal, hooked = hooked power law, LL = minus log likelihood – lower values indicate a better fit; Vuong = Vuong z values – bold if significant at p=0.05, where positive values indicate that the first bracketed model is a better fit than the second.

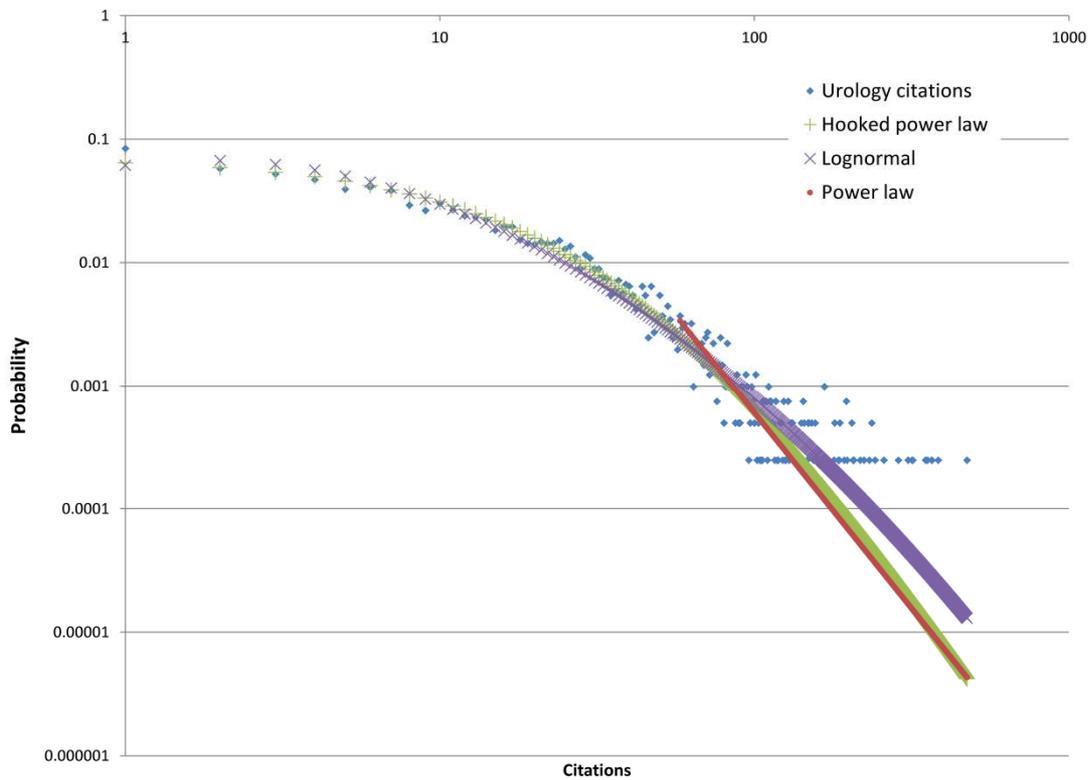

**Figure 1**. The hooked power law and the lognormal distribution fitted to the complete data set of Urology articles from 2004, together with the power law fitted to its optimal range (articles with at least 58 citations). The broomstick shape of the tail of the data in comparison to the curves of the distributions is due to the former being realised values and the latter being probability predictions. Both axes are on a logarithmic scale.

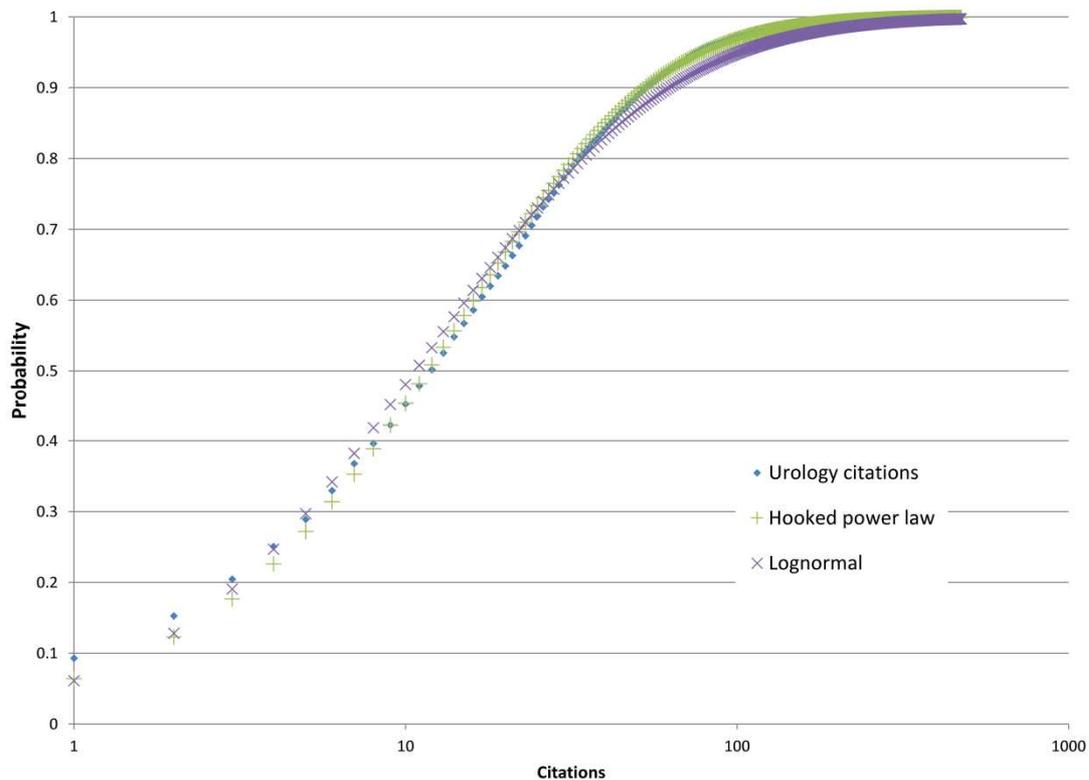

**Figure 2**. The same data as for Figure 1 except that the cumulative probabilities for the Urology data are plotted for the lognormal and hooked power law distributions, showing the closer fit for the hooked power law distribution.

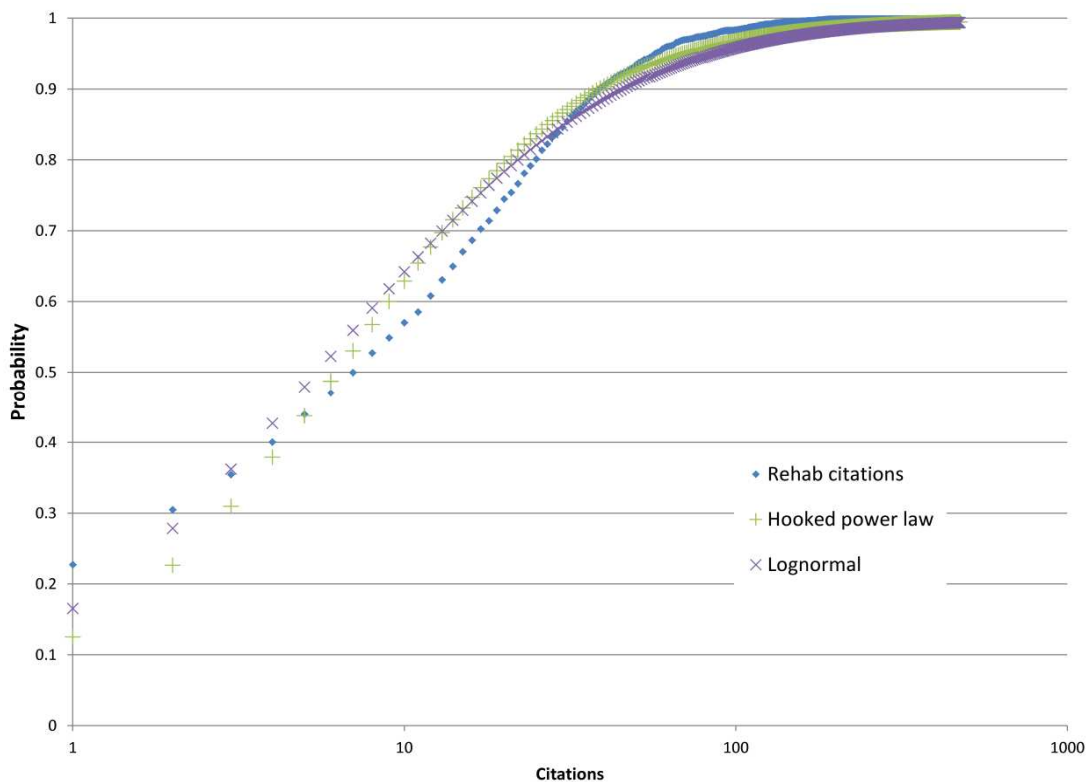

**Figure 3**. Cumulative probabilities for the Rehab data for the lognormal and hooked power law distributions, showing the closer fit for the lognormal distribution.

In answer to the second research question, the power law alpha coefficients in Table 2 are substantially lower (range: 2.238 to 4.129) than the alpha coefficients for the hooked power law (range 2.792 to 7.620). Assuming that the hooked power law is the more suitable distribution, this suggests that fitting a pure power law to the tail of the distribution is likely to give an unreliable underestimate of the exponential component (scaling parameter) of the distribution. This conclusion is supported by the fact that the hooked power law alpha coefficients in tables 3 (range 2.994 to 7.857) and 4 (range 2.514 to 10.914), which are fitted to larger sets of data and hence may be more reliable, are also substantially larger than the power law coefficients in Table 2. The reason for the problem is clear from Figure 1: there is a noticeable curve to the hooked power law for the points in which the power law is fitted (see also Appendix C).

## *4.1 Parameter estimate precision*

In answer to the third research question, the hooked power law estimates of the alpha and B coefficients for the same subject vary substantially between tables 2 to 4, showing that the sample chosen to fit the distribution has a significant impact on the results. The closest matches between tables 3 and 4 are History, Marketing and StatsProb, all of which have the same data except for excluding articles with a single citation in Table 3. Despite this small difference, the coefficients change from 2.994 and 5.0 to 2.769 and 3.7 (History), 3.465 and 39.3 to 3.437 and 38.5 (Marketing) and 3.111 and 19.9 to 3.056 and 18.8 (StatsProb). The

differences are even more marked for other cases, such as Ecology (7.857 and 143.4 to 10.914 and 233.7). Hence it seems unlikely that the hooked power law parameter estimates in Table 4 are within, say, 10% of the true value, assuming that the data is taken exactly from a hooked power law. The same is true for the lognormal distribution. For example, the coefficients change from 0.8 and 1.29 to 0.6 and 1.36 (History), 2.6 and 1.25 to 2.5 and 1.32 (Marketing) and 2.1 and 1.26 to 2.0 and 1.33 (StatsProb). Although confidence intervals for the parameters in tables 2 to 4 could have been estimated using a bootstrapping approach, the results may have been misleading since the high citation counts are important for the distribution and these could not be dealt with effectively with bootstrapping. Instead, the next subsections discuss the accuracy of parameter estimates in general, using simulated data.

### 4.1.1 Hooked power law parameter estimates

In order to systematically assess the accuracy of the estimates of the key parameter of the hooked power law distribution, simulations were run at different sample sizes of about the magnitude of a single year of citation data. For each value of alpha in the range 2 to 10 and for each sample sizes of 500, 1000, 2000, and 4000, a distribution was simulated 500 times by selecting data points at random from the distribution for a range of alphas and B = 10. From each set of 500 simulations a 90% confidence interval was constructed for alpha and the width of the confidence interval was measured. From the results in Figure 4, it is clear that the alpha estimates are less variable at higher sample sizes, which is unsurprising, but also that for high values of alpha the estimates are extremely variable. The smallest confidence interval width in the graph is 0.12, which is for alpha=2 and a sample size of 4000 and so it is clear that sample sizes of above 4000 are needed to get one decimal place accuracy, even for low values of alpha and for data that perfectly fits the distribution.

      The relative sizes of the confidence intervals can also be seen from the fact that the range of the alpha values in Table 4 is 8.4 and so the confidence intervals in Figure 4 for a sample of size 500 vary from 5% to about 45% of this range.

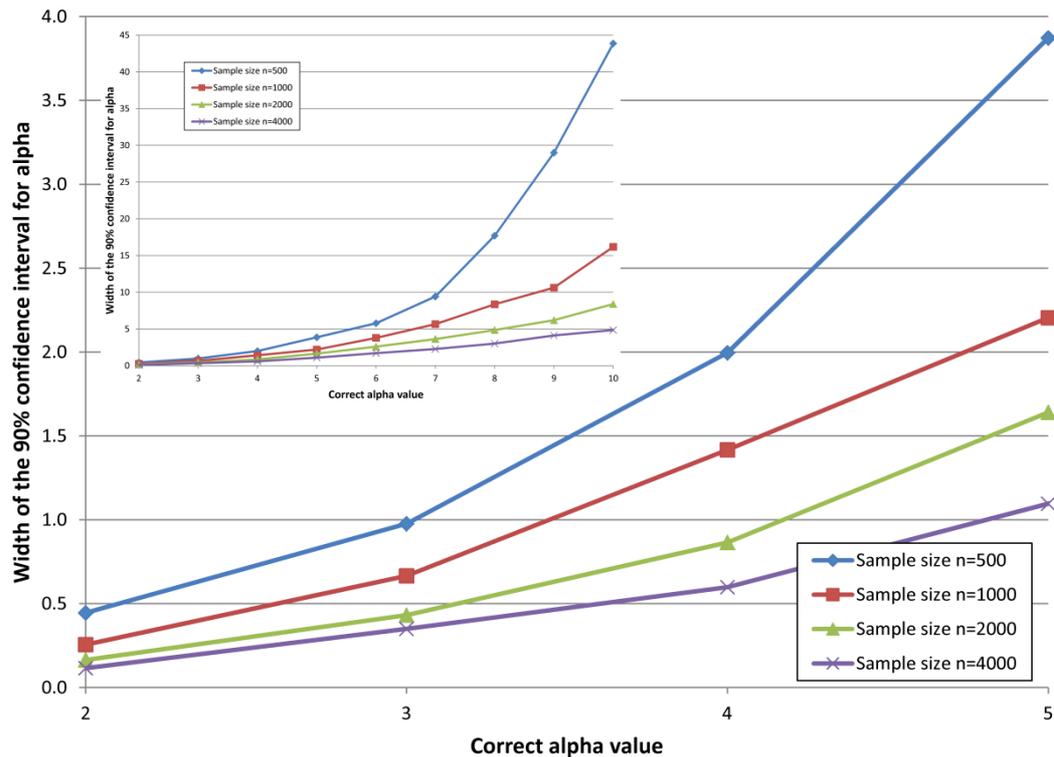

**Figure 4**. Confidence interval widths for simulated hooked power laws at different values of alpha, with B=10, and for different sample sizes (500 simulations per data point).

The reason for the poor precision of the alpha estimates and also the $B$ estimates (not shown) is that the alpha and $B$ values can offset each other to some extent on the sampled data. To illustrate this, Figure 5 shows data randomly sampled from a distribution with alpha=3 and $B = 2$, where the best fit to the randomly sampled data is alpha = 4.8 and $B = 24.3$. The closeness of fit to the data of both the correct distribution and the best fit distribution is clear in Figure 5. If the fitted alpha is combined with the correct $B$ value, however, the fit is very poor, suggesting that the increase in alpha in the best fit solution has been compensated for by the increase in B. To test for this, a contour plot of log-likelihoods for different values of alpha and $B$ was plotted (Figure 6). The linear shape in Figure 6, which is replicated for the other samples (not shown) confirms that increasing alpha can be compensated for by increasing $B$.

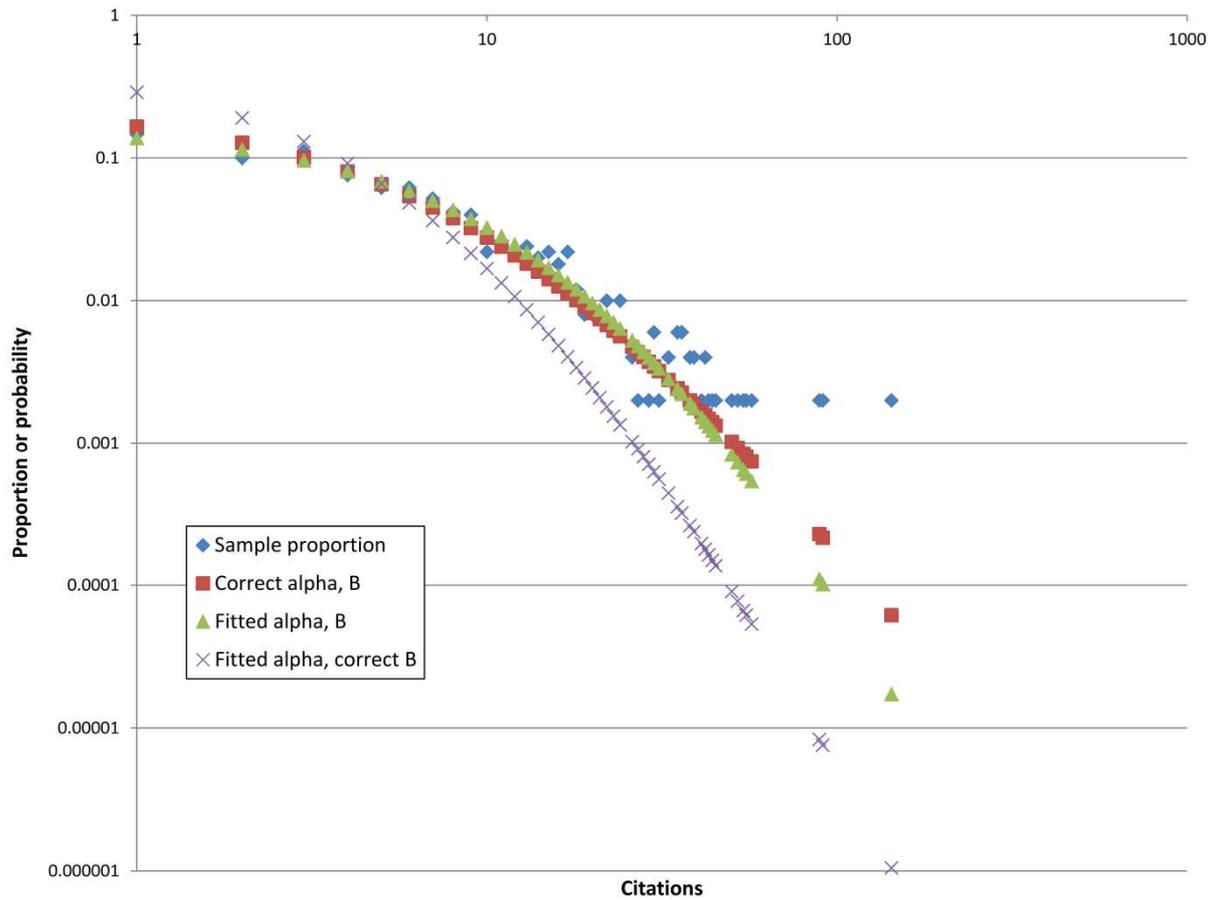

**Figure 5**. A set of 500 sample citations taken from a hooked power law distribution with alpha=3 and B=10, together with corresponding figures from the original (correct) distribution, the hooked power law best fitting the sample data (alpha = 4.8 and $B = 24.3$) and a hybrid distribution (alpha = 4.8 and $B = 10$).

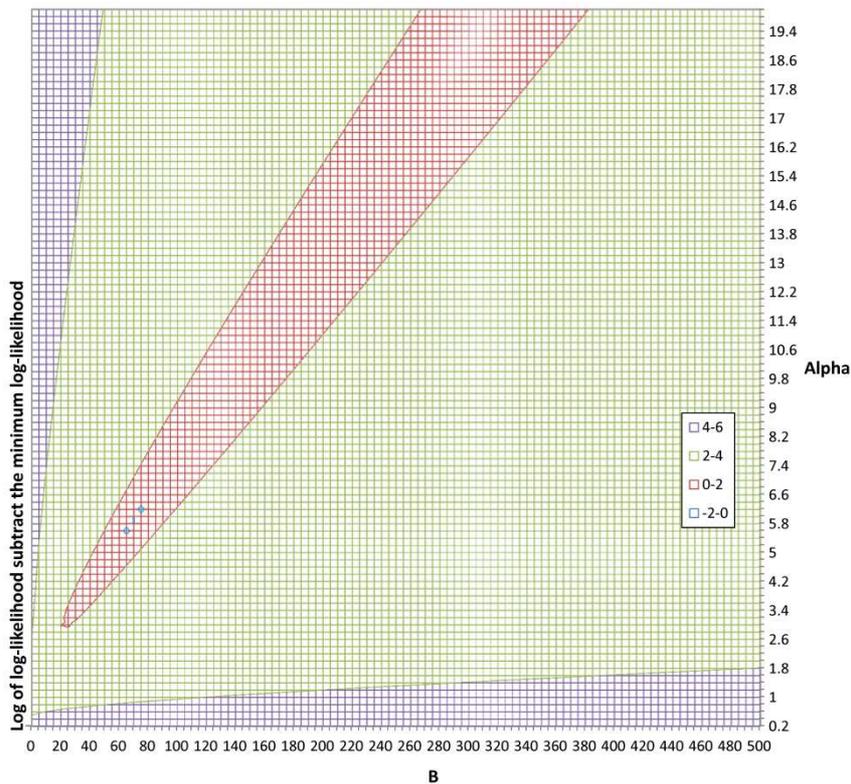

**Figure 6**. A contour plot of log-likelihoods for different hooked power law parameters for the Soil data. Lower values indicate a better fit.

### 4.1.2 Lognormal parameter estimates

The widths of the 90% confidence intervals for the mean and standard deviation of the lognormal distribution based on 500 simulations of random sampling from pure distributions are much narrower than the 90% confidence intervals for alpha in the hooked power law (figures 7-10 – a surface plots are used rather than a line graph similar to Figure 4 because the two parameters behave differently), suggesting that the parameters in the lognormal distribution can be fitted much more precisely than can hooked power law distributions, at least in absolute terms. The confidence intervals are smallest, and hence the precision highest, when the mean is larger and the standard deviation is smaller. When the sample size is 500, the mean and standard deviation are likely to have a precision of zero decimal places (figures 7, 9), and so there should be little confidence about the first decimal place. When the sample size is 4000, the mean is likely to have a precision of about 0.25, but one decimal place in the case of high means and low standard deviations (Figure 8). At this sample size, the standard deviation is likely to have a precision of one decimal place for many combinations of mean and standard deviation but for other values, the first decimal place could be out by 0.1 or 0.2 (Figure 10).

The relative sizes of the confidence intervals can be seen from the fact that the range of the means in Table 4 is 2.8 and so the mean confidence intervals in Figure 7 for a sample of size 500 vary from about 4% to about 25% of this range. Similarly, the range of the standard deviations in Table 4 is 0.68 and so the mean confidence intervals in Figure 9 for a sample of size 500 vary from about 6% to about 41% of this range.

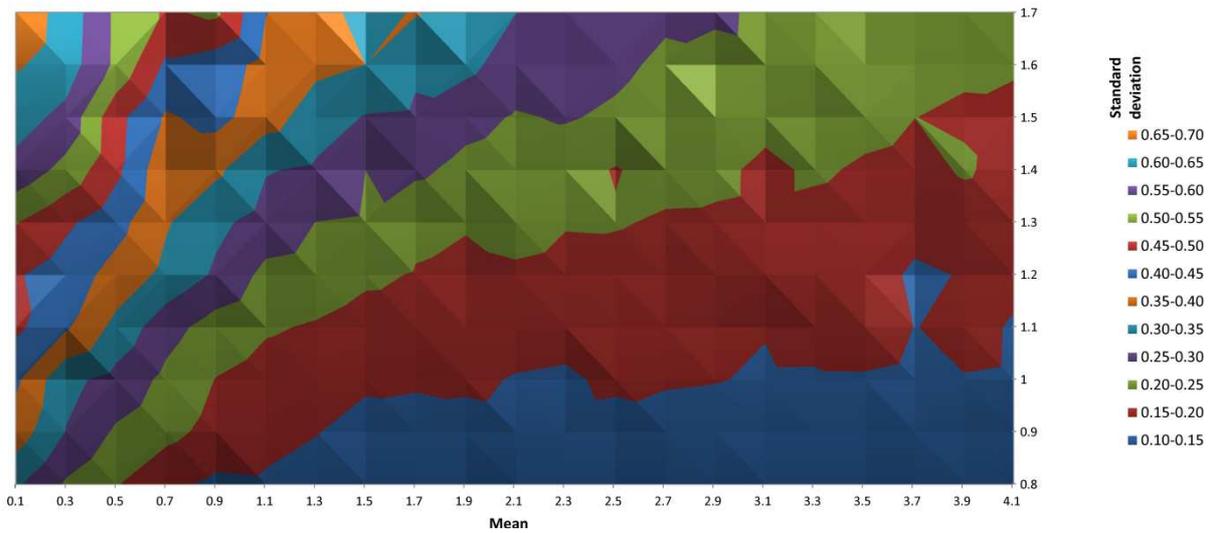

**Figure 7**. A contour plot of 90% confidence interval widths for the *mean* of the lognormal distribution for different values of the mean and standard deviation and a sample size of *500* (500 simulations per data point).

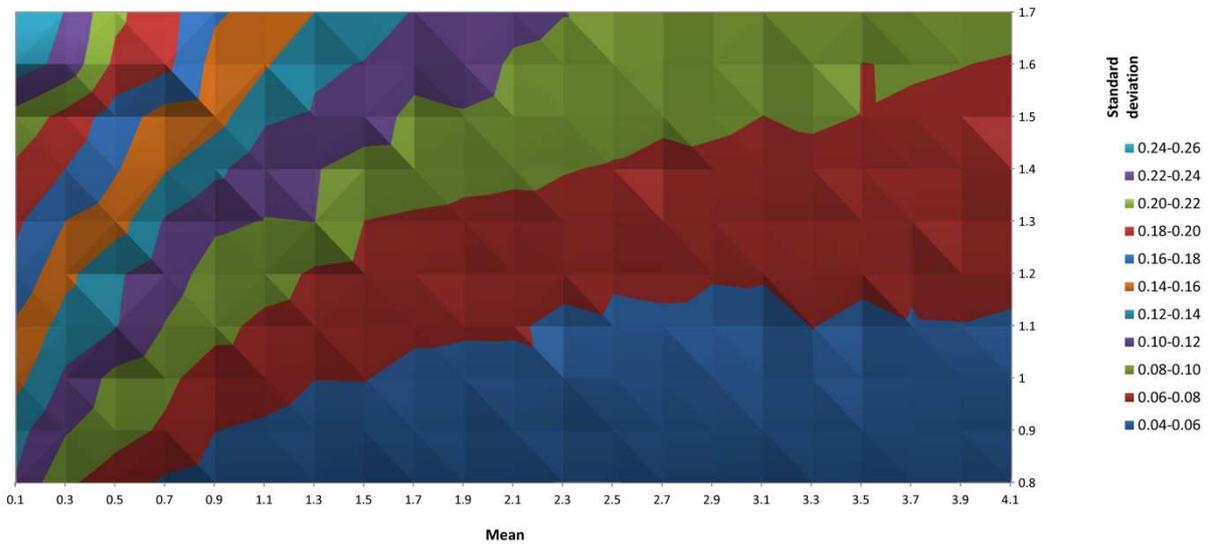

**Figure 8**. A contour plot of 90% confidence interval widths for the *mean* of the lognormal distribution for different values of the mean and standard deviation, and a sample size of *4000* (500 simulations per data point).

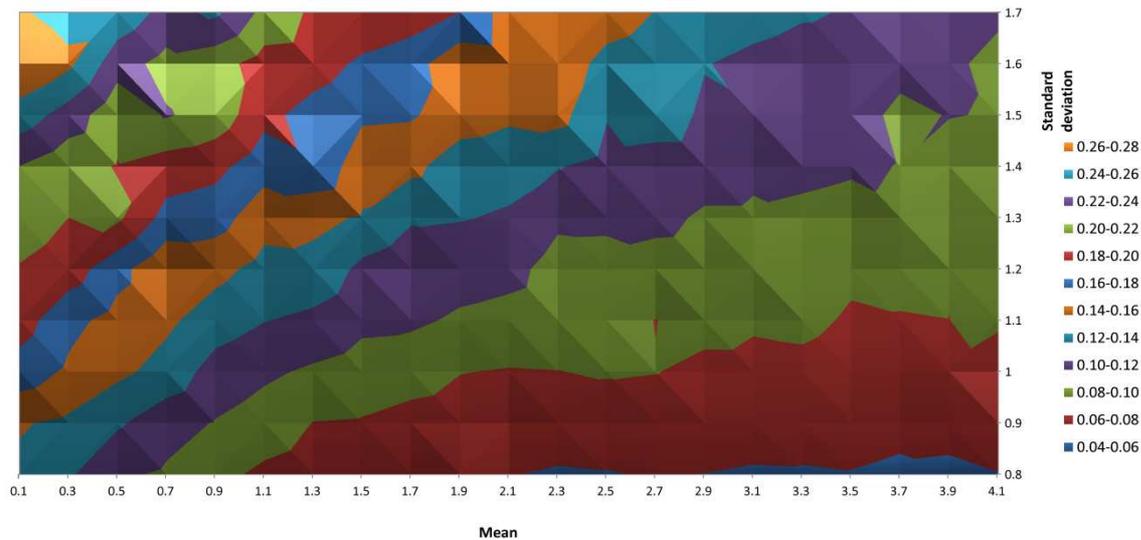

**Figure 9**. A contour plot of 90% confidence interval widths for the *standard deviation* of the lognormal distribution for different values of the mean and standard deviation and a sample size of *500* (500 simulations per data point).

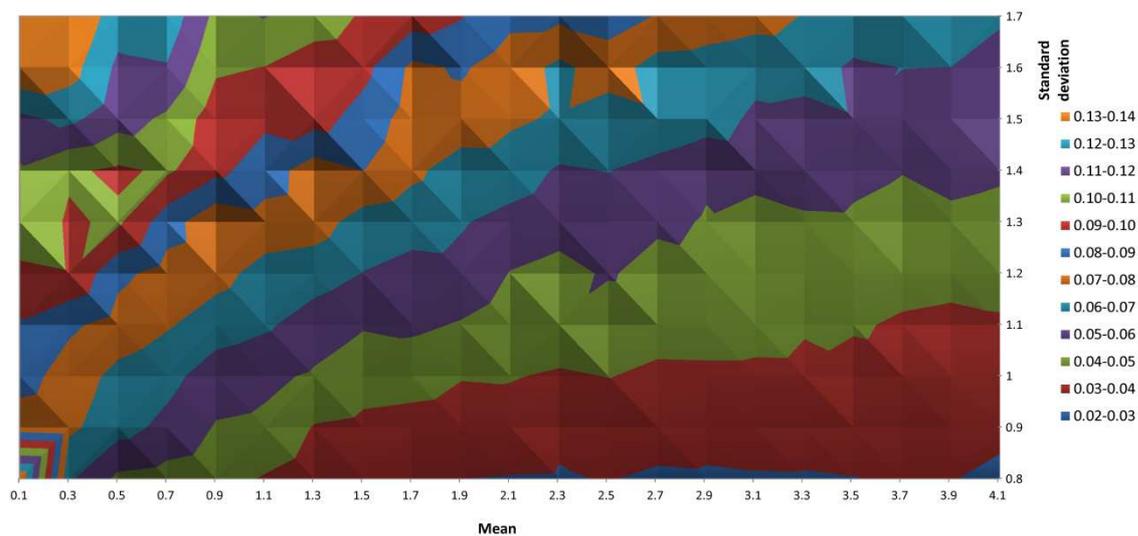

**Figure 10**. A contour plot of 90% confidence interval widths for the *standard deviation* of the lognormal distribution for different values of the mean and standard deviation and a sample size of *4000* (500 simulations per data point).

Figure 11 is a contour plot for the log-likelihood of the lognormal distribution for different values of the mean and standard deviation for the Soil data set. In contrast to the similar contour plot for the hooked power law (Figure 6), there is a clear focal point (mean: 2.3; deviation: 1.2) and it is not possible to compensate for increases in the mean by changing the standard deviation, and vice versa. Hence, the values of parameters in fits of the lognormal distribution should intuitively be more precise than the values of parameters in fits of the hooked power law, in the sense of being less subject to random fluctuations in the samples analysed, confirming the discussion above.

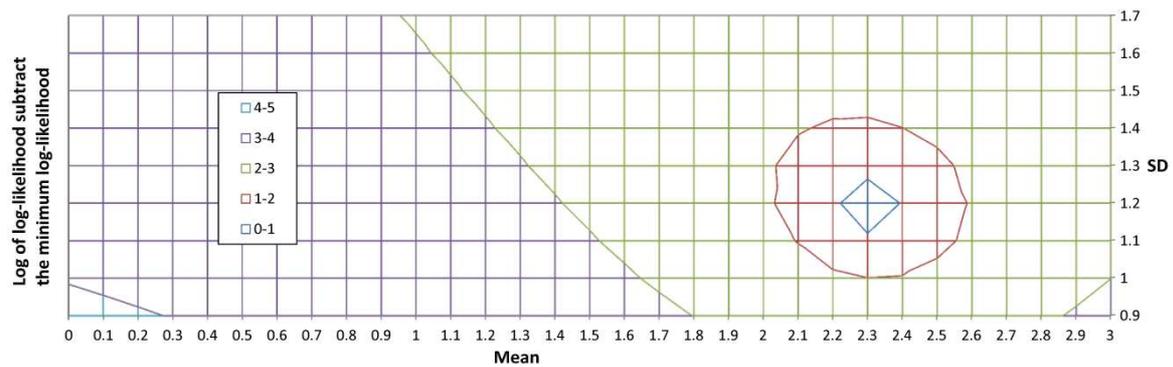
**Figure 11**. A contour plot of log-likelihoods for different lognormal parameters for the Soil data. Lower values indicate a better fit.

# 5. Conclusions

The results suggest that fitting a power law to citation data on the scale here (up to 5000 papers) is not appropriate and has no use. Even for the tail of the data for which the power law fits best, the hooked power law and lognormal distributions probably fit marginally better (even accounting for the benefits of their extra parameter, as the Vuong test does) and they both fit substantially better when more of the tail is included. Moreover, the exponent alpha estimated from the power law from the optimal tail size is typically a substantial underestimate of the exponent alpha from the hooked power law and so even this estimate has no value. Most previous studies of citation distributions have focused only on modelling the tail, however, and these findings may not apply in this context.

The results also suggest that the hooked power law and lognormal distributions are approximately equivalent in their fit to citation data of the type analysed (a maximum of 5000 articles from one subject and one year and about 10 years of citation data). Although the hooked power law fits better than the lognormal distribution for the full data sets (for articles with at least one citation) more often than the other way around, both fit statistically significantly better than each other for some subjects and so it is not possible to claim that one is universally the best fit. A possible explanation is that the data has anomalies at low citation counts (e.g., caused by low impact special issues of journals) which obscure the underlying distribution. Ultimately, however, all statistical distributions will not fully account for all factors associating with citations and hence both the hooked power law and lognormal distributions can only be approximate fits.

The clear advantage of the hooked power law and lognormal distributions over the power law suggests that either would be reasonable to use for modelling single subjects and years. Nevertheless, the unreliability of the parameter estimates for both distributions, but particularly for the hooked power law, suggests that their values cannot be calculated with a high precision and, therefore, that any attempt to compare two citation distributions (e.g., biochemistry vs. chemistry or USA vs China) to assess whether they are the same would need to take this into account. The slightly greater accuracy with which the lognormal distribution can be fitted makes it the distribution of choice for analyses of citations to academic articles for a single subject and year, however.

## Appendix A: The hooked power law for citations

Translating the original hyperlink context (Pennock et al., 2002) to citations, the hooked power law assumes that the probability of a new reference pointing to a previously-published page $i$ is:

$$\beta \frac{k_i}{2mt} + (1-\beta)\frac{1}{m_0 + t}$$

Here $\beta$ is the probability that a cited paper is chosen on the basis of the rich-get-richer mechanism underlying the power law (the probability that a paper receives a new citation is proportional to the number of citations that it already has) and $1 - \beta$ is the probability that the cited paper is chosen at random. In the formula, $t$ (=1,2,…) is the time when the $t$th new paper is added to the system, $m_0$ is the number of papers when the process starts, $m$ is the (fixed) number of references (to other papers in the system) in each new paper, and $k_i$ is the number of references already received by the paper (i.e., its current citation count).

This formulation is an oversimplification of the situation for citations. Although, as discussed above, it plausible that papers sometimes get cited because of their citations and so the first half of the calculation seems possible, the second half does not. Presumably articles are almost never cited purely at random, even if there is a degree of arbitrariness about the selection of articles to cite, and so the right hand side of the expression should take into account the fact that the quality or utility of articles is not equal so a purely random selection process is unrealistic. The problems caused by this oversimplification may tend to even out if enough data is collected, but this cannot be assumed. A necessary (but not sufficient) condition for this would be to assess whether the assumption leads to a distribution that approximately fits real-world data.

If many papers are added to a system in a way that obeys the above rule then the probability that a random paper has received $k$ citations is (Pennock et al., 2002):

$$P(k) = [2m(1-\beta)]^{1/\beta}[\beta k + 2m(1-\beta)]^{-1-1/\beta}$$

At a fixed point in time, the only variable in the above formula is k, the citation count, and so the above simplifies to the hooked power law formulation for the probability of an article receiving $k$ citations $A/(B + k)^\alpha$, where $A = [2m(1-\beta)]^{1/\beta}\beta^{-1-1/\beta}$, $B = 2m(1-\beta)/\beta$ and $\alpha = 1 + 1/\beta$. The simpler form is used in the remainder of this article.

# Appendix B: Fitting truncated distributions

A probability distribution for articles with at least $x_{min}$ citations is a formula $p(x)$ for the probability that an article has x citations. It must have the properties that $0 \leq p(c) \leq 1$ for all positive integers $x$ and $\sum_{x=x_{min}}^{\infty} p(x) = 1$.

A hooked power law probability distribution has two parameters, $\alpha$ and $B$, and is given by $p(x) = A/(B+x)^{\alpha}$, where A is a constant chosen to ensure that $\sum_{x=x_{min}}^{\infty} p(x) = 1$. If $B = 0$ then this is a simple power law rather than a hooked power law. The constant A is equal to $1/\sum_{x=x_{min}}^{\infty} 1/(B+x)^{\alpha}$. This is a formula without a simple analytical solution. For example, if B=0 and $x_{min} = 1$ then it simplifies to the Riemann zeta function and if B>0 and $x_{min} = 0$ and $\alpha > 1$ then it coincides with the Hurwitz zeta function. When needed, these values were approximated by taking the sum of the first 10,000 terms of the series. Although this approximation is not recommended, its accuracy was checked by comparing the results with known values of the Riemann zeta function, such as $\pi^2/6$ for B=0 and $\alpha = 2$. The same procedure (summing the first 10,000 terms) terms was also used to calculate normalising constants for the lognormal distribution, when needed (e.g., Figure 1 and the Voung tests).

The power law and lognormal distributions were fitted to the data using the poweRlaw R package, which implements the procedures described by Clauset, Shalizi, and Newman (2009). The hooked power law was fitted using a numerical variant of the classical gradient descent method, implemented in R. Although the gradient descent problem, in general, can get trapped in local minima, testing on the data sets suggested that this was not likely to happen when fitting the hooked power law to large data sets for which is it is a reasonable fit. Vuong tests for the fit between the power law and lognormal distribution were taken from R and Vuong tests for the fit between the hooked power law and the lognormal were implemented in R, following Voung's (1989) formula.

# Appendix C: Poor estimates of alpha from truncated power laws

In order to demonstrate that power law exponents cannot be estimated from fitting truncated power laws unless only papers with a very high numbers of citations are used, assume that the hooked power law is the correct distribution for a set of citations and consider the graph of the hooked power law $y = A/(B+x)^{\alpha}$ on the logarithmic scale, where the estimate for $-\alpha$ essentially depends on the slope of the curve at a point. Let X=log(x) and Y=log(y) so that the graph of Y against X on standard axes would be the same as the graph of y against x on double logarithmic scale, with formula $Y = \log(y) = \log\left(\frac{A}{(B+x)^{\alpha}}\right) = logA - \alpha \log(B+x) = logA - \alpha \log(B + e^X)$. Now the derivative of Y is $Y' = \frac{-\alpha e^x}{B+e^x}$ and this is the slope of the graph and hence is used to estimate alpha. Assume that this gradient must be close to alpha, say within a tolerance of $T\alpha$. For example if T=0.01 then the gradient (slope) would have to be within 1% of the true value of alpha. Now at this tolerance $Y' = -\alpha + T\alpha$ and substituting the derivative above, $\frac{-\alpha e^x}{B+e^X} = T\alpha - \alpha$. Solving this equation for X gives $e^X = \frac{1-T}{T}B$ or $x = \frac{1-T}{T}B$. Hence, for example, the slope of the hooked power law graph starts to be within 10% of the value of alpha when x, the number of citations, equals 9B. So if B=55 (the average value in Table 4) then the slope would get within 10% of the value of alpha only for articles with above 9x55=495 citations, which is much too high to be practical as an $x_{min}$ value, but with smaller values of $B$ then the alpha estimate would likely be too low by about 5% (assuming averaging over the interval initially 10% too steep to exactly the right slope).